\documentclass[twocolumn,english,superscriptaddress,secnumarabic,amssymb,amsmath,nobibnotes,aps,prd,showkeys,showpacs]{revtex4}
\usepackage[latin9]{inputenc}
\usepackage{color}
\usepackage{verbatim}
\usepackage{textcomp}
\usepackage{amsmath}
\usepackage{amssymb}
\usepackage{graphicx}
\usepackage{esint}
\usepackage{ulem}
 
\newcommand{\pd}[2]{\frac{\partial #1}{\partial #2}} 
 
\let\baraccent=\= 
\renewcommand{\=}[1]{\stackrel{#1}{=}} 

\begin{document}

 \title{The dynamics of a spinning particle in a linear in spin Hamiltonian
 approximation}

\author{Georgios Lukes-Gerakopoulos}
\email{gglukes@gmail.com}
\affiliation{Institute of Theoretical Physics, Faculty of Mathematics and Physics, 
 Charles University in Prague, Czech Republic}
\author{Matthaios Katsanikas}
\email{mkatsan@academyofathens.gr}
\affiliation{Research Center for Astronomy, Academy of Athens, Soranou Efessiou 4, GR-115 27, Athens, Greece}
\author{Panos A. Patsis}
\email{patsis@academyofathens.gr}
\affiliation{Research Center for Astronomy, Academy of Athens, Soranou Efessiou 4, GR-115 27, Athens, Greece}
\author{Jonathan Seyrich}
\email{seyrich@na.uni-tuebingen.de}
\affiliation{Mathematisches Institut, Universit\"{a}t T\"{u}bingen,
Auf der Morgenstelle, 72076 T\"{u}bingen, Germany}

\begin{abstract}
 We investigate for order and chaos the dynamical system of a spinning test
 particle of mass $m$ moving in the spacetime background of a Kerr black
 hole of mass $M$. This system is approximated in our investigation by
 the linear 
 in spin Hamiltonian function provided in {[}E. Barausse, and A. Buonanno,
 \textit{Phys.Rev. D} \textbf{ 81}, 084024 (2010){]}. We study the corresponding
 phase space by using 2D~projections on a surface of section and the method of
 color and rotation on a 4D~Poincar\'{e} section. Various topological structures
 coming from the non-integrability of the linear in spin Hamiltonian are found
 and discussed. Moreover, an interesting result is that from the
 value of the dimensionless spin $S/(m~M)=10^{-4}$ of the particle
 and below, the impact of the non-integrability of the system on the motion
 of the particle seems to be negligible.
 
\end{abstract}

%
%
\maketitle

\section{Introduction}\label{sec:Intro}

 Mathisson \cite{Mathisson37} and Papapetrou \cite{Papapetrou51} provided the
 equations of motion for a spinning particle in a curved spacetime. The
 equations of motion of a spinning test particle are interesting from the
 astrophysical point of view, because they approximate the motion of a stellar
 compact object in the spacetime background of a supermassive black
 hole. Such a binary 
 system is called an extreme mass ratio inspiral (EMRI). EMRIs are
 among the most 
 promising sources of gravitational waves expected to be detected by space
 interferometer antennas like LISA (see, e.g., \cite{LISA}). However, in this
 work we focus rather on the dynamics of a spinning particle system than on 
 astrophysical aspects.
 
 The number of Mathisson-Papapetrou (MP) equations is smaller than the number of
 variables which the MP equations intend to evolve. The above fact can
 be interpreted 
 as a freedom of choosing different worldlines for evolving the equation of
 motion of the same extended object described by the pole-dipole approximation
 \cite{Moeller49}. To choose a worldline we use a supplementary condition that
 is known in the literature as the spin supplementary condition (SSC). There is
 a variety of SSCs (for a review, see, e.g., \cite{Semerak99,Kyrian07,Semerak15}),
 but all are physically acceptable. The most renown are the Pirani (P) 
 \cite{Pirani56} and the Tulczyjew (T) \cite{Tulczyjew59} SSCs. For many years 
 the P SSC was considered unphysical, because the test particle exhibited helical
 motion in the flat spacetime limit. However, in \cite{Costa12} it has been shown
 that this helical motion results from a hidden momentum and the P SSC is 
 physically valid as well.
 
 The aspect of the spinning particle dynamics we are interested in is the issue
 of the integrability of the corresponding system. It has been shown that for 
 the Schwarzschild background the MP equations with T SSC give chaotic orbits
 \cite{Suzuki97}, and the same holds for the Kerr background, see, e.g.,
 \cite{Hartl03a,Hartl03b,Han08}. Hence, one can claim that the MP equations with
 T SSC  correspond to a non-integrable system. However, in the linear in spin
 approximation of the MP equations, it has been proved that for the T SSC a
 Killing-Yano tensor provides a Carter-like constant of motion for the Kerr
 background \cite{Rudiger}. The existence of a Carter-like constant and the fact
 that the T and P SSCs are the same in the linear regime led to the impression 
 that in the linear in spin approximation the spinning particle dynamics 
 corresponds to an integrable system (see, e.g., \cite{Hinderer13}). For 
 geodesic orbits in a Kerr spacetime the existence of the Carter constant
 ensures integrability, since it is the fourth constant of motion (the others
 being the energy, the angular momentum along the symmetry axis and the
 contraction of the four-momentum) in a Hamiltonian system of four degrees of 
 freedom. Nevertheless, when the particle is spinning we have extra degrees of
 freedom, and it is questionable even if the existence of a Carter-like constant
 can ensure the integrability of the system.  
 
 When examining whether a system is integrable or not, it is useful to have
 a canonical Hamiltonian formalism which provides symplecticity. In a
 non-symplectic system we need as many constants of motion as the dimensionality
 of  the phase space. On the other hand, in a canonical Hamiltonian system two 
 dimensions of the phase space correspond to one degree of freedom. Therefore,
 we need half the number of constants of motion in order to have integrability 
 with respect to a non-canonical system of the same phase space
 dimensionality. Moreover, by having a canonical Hamiltonian system, tools like
 Poincar\'{e} sections can be properly used. When a system is not symplectic,
 then a surface of section is  ambiguous. A canonical Hamiltonian formalism has
 not been found yet for the MP equations with the T SSC. However, such canonical
 Hamiltonian formalism has been provided in \cite{Barausse09} for the
 Newton-Wigner (NW) SSC \cite{NewtonWigner49} in the linearized in spin
 approximation. 
 
 A Hamiltonian for a spinning particle moving in a Kerr spacetime background 
 has been first provided in \cite{Barausse09}. However, due to 
 the approximative procedure which leads from the MP equations to the linearized
 in spin Hamiltonian function, the resulting Hamiltonian equations are not
 equivalent to the corresponding MP equations, e.g., starting the Hamiltonian
 equations and the MP equations with the same initial conditions lead to two
 diferrent orbits \cite{LGSK}. It might even occur that the final linearized
 Hamiltonian system may not even respect some symmetries that the corresponding
 MP equations respect. For example, the Hamiltonian function provided
 in \cite{Barausse09} for the Kerr spacetime in Boyer-Lindquist coordinates did
 not respect in the Schwarzschild limit $a=0$ the spherical background symmetry.
 Namely, the total angular momentum was not preserved as it should be. Absence
 of integrals of motions could lead to the misleading impression that for the
 Schwarzschild background the Hamiltonian corresponds to a non-integrable system
 (see, e.g., figure~2 in \cite{KLLS}). The problem with the Hamiltonian in 
 \cite{Barausse09} was the specific tetrad field choice on which the Hamiltonian 
 function was built on. Even in \cite{Barausse09}, it was found that the resulting
 Hamiltonian would evolve the spin in the flat spacetime limit \cite{Barausse09}.
 However, since the helical motion in the case of the MP equations with P SSC
 could result from a hidden momentum, the same could hold for the Hamiltonian
 approximation coming from the NW SSC. Thus, a more solid reasoning was needed
 to show the drawbacks of the tetrad field chosen in \cite{Barausse09}. In
 \cite{KLLS} it was shown that if the resulting Hamiltonian should respect 
 the symmetries of the Schwarzschild background, then the corresponding tetrad
 filed should obey a certain prescription (equation~(44) in \cite{KLLS}). The
 tetrad field of \cite{Barausse09} is not complying with this prescription.
 
 On the other hand, a different tetrad field choise provided in
 \cite{Barausse10} led to a revised Hamiltonian for the Kerr background. This
 tetrad field is obeying the prescription given in \cite{KLLS}. In particular,
 for the Schwarzschild limit the revised Hamiltonian \cite{Barausse10} was
 conserving not only the total angular momentum as it should be, but it was
 shown in \cite{KLLS} that the magnitude of the orbital momentum was preserved
 as well. The latter implies that in the Schwarzschild limit the revised
 Hamiltonian of \cite{Barausse10} corresponds to an integrable system, since
 for a five degrees of freedom system we have five constants of motion
 \cite{KLLS}\footnote{In fact it was shown that a proper Hamiltonian function
 for the Schwarzschild background corresponds to an integrable system in general.}. 
 The revised Hamiltonian ceases to be integrable when the spin of the central
 black hole is nonzero \cite{KLLS}, i.e., in the Kerr spacetime background. A
 thorough study of the non-integrability of the revised Hamiltonian is the
 subject of our article.

 The study of chaotic motion around black holes probably starts
 with \cite{Contopoulos90}, where a method based on Cantor sets was applied to 
 prove the chaotic nature of the system. Since then many methods have been
 applied to detect chaos in the vicinity of black holes, but the most common is
 the 2D Poincar\'{e} section. It is a fact that in order to study
 the non-integrability of a two degrees of freedom Hamiltonian system a 2D
 Poincar\'{e} section is a standard tool. However, since the Hamiltonian
 provided in \cite{Barausse10} corresponds to a system with three degrees of
 freedom, we have to deal with 4D Poincar\'{e} sections \cite{Contopoulos02}.
 In order to detect order and chaos in the 4D Poincar\'{e} spaces of section,
 we must first of all have a way to visualize them. In the past, several methods
 have been proposed for the visualization of the 4D surfaces of section in a 6D
 phase space of a 3D autonomus Hamiltonian system: ordinary 2D projections
 \cite{Contopoulos89}, 3D projections \cite{Vrahatis97}, stereoscopic projections
 \cite{Froeschle70,Martinet81,Contopoulos82}, or 2D slices of 3D
 subspaces (\cite{Froeschle70,Froeschle72}, and recently a more sophisticated
 version in \cite{Richter14,Lange14} (see Appendix).

 In the present work we use the method of color and rotation, introduced by
 Patsis and Zachilas \cite{Patsis94}. This method is extensively described for
 the case of 3D rotating galactic potentials in a series of papers
 \cite{Katsanikas11a,Katsanikas11b,Katsanikas11c,Katsanikas13,Patsis14a,Patsis14b}.
 These papers investigate portraits of the 4D spaces of section in the
 neighborhood of periodic orbits exhibiting all kinds of instabilities
 encountered in 3D Hamiltonian systems (see, e.g., \cite{Contopoulos02}). The
 method has also been applied in the study of the structure of the phase space
 close to fixed points in a 4D symplectic map \cite{Zachilas13}, and to design
 spacecraft orbits \cite{Geisel13}.

 The method consists in plotting the points (the consequents) of an orbit in a
 3D subspace as they cross the space of section in a given direction, rotate 
 them by means of standard 3D graphic tools to get a good insight of their
 distribution in the 3D subspace, and finally color them according to their value
 in the fourth dimension (the one not used in the 3D spatial representation of 
 the orbit). Color allows the estimation of the smoothness in the 4th dimension
 of geometrical structures appearing in the 3D projections and the distinction
 of pseudo- from true intersections in the 4D space. Thus, one can establish 
 criteria for the regular, weak chaotic or strong chaotic character of a given orbit 
 \cite{Patsis94,Katsanikas11a,Katsanikas11b,Katsanikas11c,Katsanikas13,Patsis14a,Patsis14b,Zachilas13}.
 In the latter papers, specific patterns in phase space are associated with the 
 various kinds of instability or with stability. In our paper this method is
 used in the study of the dynamics of a spinning particle in the Hamiltonian
 approach in an effort to trace regular and chaotic motion in the phase
 space of our system.
 
 The paper is organized as follows. Sec.~\ref{sec:HamSP} introduces the 
 Hamiltonian function of \cite{Barausse10}, which we use for our study.
 Sec.~\ref{sec:2D4DPs} discusses the non-integrability of the Hamiltonian,
 briefly describes the setting up of the numerics, and provides a detailed 
 account of our numerical findings. Sec.~\ref{sec:ConDis} sums up our findings,
 and discusses the possible astrophysical implications. Appendix~\ref{sec:V4D}
 lists techniques used for visualizing 4D spaces of section.
 
 We use geometric units, i.e., $G=c=1$, and the signature of the metric is
 (-,+,+,+). Greek letters denote the indices corresponding to spacetime (running
 from 0 to 3), while Latin letters denote indices corresponding only to space 
 (running from 1 to 3).

\section{The Hamiltonian of a spinning particle} \label{sec:HamSP}

 The canonical Hamiltonian formalism of a spinning particle in \cite{Barausse09}
 has been achieved by linearizing the MP equations of motion for the NW SSC. 
 In this formalism the mass of the test particle $m$ is considered a constant of
 motion \cite{Barausse09}, and the spin of the particle is given by a three
 vector $S^I$. The corresponding Hamiltonian function $H$ splits in two main parts,
 the non-spinning $H_{NS}$, which describes basically the geodesic motion, and
 the spinning part $H_S$,  which incorporates the spinning of the particle, i.e.,
 \begin{equation}
  H=H_{NS}+H_{S}~~. \label{eq:HamSP}
 \end{equation}

 The non-spinning part of the Hamiltonian $H_{NS}$ reads
 \begin{equation}
  H_{NS}=\beta^{i}P_i+\alpha~\sqrt{m^2+\gamma^{ij}P_i P_j}~~, \label{eq:HamNSP}
 \end{equation}
 where $P_i$ are the canonical momenta conjugate to the coordinates $x^{i}$ of the
 Hamiltonian~\eqref{eq:HamSP} \cite{Barausse09}, and
 \begin{eqnarray}
  \alpha &=& \frac{1}{\sqrt{-g^{00}}}~~, \nonumber  \\
  \beta^i &=& \frac{g^{0i}}{g^{00}}~~, \nonumber \\
  \gamma^{ij} &=& g^{ij}- \frac{g^{0i}g^{0j}}{g^{00}}~~. \label{eq:abg}
 \end{eqnarray}
 $g^{\kappa\lambda}$ is the contravariant form of the metric tensor of the
 background spacetime in which the test particle moves. We are interested in the
 Kerr spacetime background describing the spacetime around a black hole of mass
 $M$ with spin parameter $a$. In Boyer-Lindquist coordinates $t$ is the
 coordinate time, $\phi$ is the azimuthal angle, $\theta$ is the  polar angle,
 and $r$ is the radial distance, and the Kerr metric reads
 \begin{eqnarray}
   g_{tt} &=&-1+\frac{2 M r}{\Sigma}~~,\nonumber\\ 
   g_{t\phi} &=& -\frac{2 a M r \sin^2{\theta}}{\Sigma}~~,\nonumber\\
   g_{\phi\phi} &=& \frac{\Lambda \sin^2{\theta}}{\Sigma}~~,\nonumber  \\
   g_{rr} &=& \frac{\Sigma}{\Delta}~~,\nonumber\\
   g_{\theta\theta} &=& \Sigma~~,\label{eq:KerrMetric}
 \end{eqnarray} 
 where
 \begin{eqnarray}
  \Sigma &=& r^2+ a^2 \cos^2{\theta}~~,\nonumber\\
  \Delta &=& \varpi^2-2 M r~~,\nonumber \\ 
  \varpi^2 &=& r^2+a^2~~, \nonumber \\ 
  \Lambda &=& \varpi^4-a^2\Delta \sin^2\theta~~.  \label{eq:Kerrfunc} 
 \end{eqnarray}

 The spinning part of the Hamiltonian $H_S$ for the Kerr spacetime in Boyer-Lindquist
 coordinates as given in \cite{Barausse10} can be split in two parts as well, i.e.,
 \begin{equation} \label{eq:RHamBL}
  H_S=H_{SO}+H_{SS}~~,
 \end{equation}
 where the Hamiltonian providing the spin orbit coupling reads 
 \begin{widetext}
 \begin{eqnarray}\label{eq:RHamSOBL}
  H_{SO} &=& \frac{\sqrt{\Delta~\Sigma}~P_\phi~S_z}{m \Lambda \sqrt{Q} \sin^2{\theta}}
  (\frac{\Sigma}{\sqrt{\Lambda}}-1)+\frac{1}{\sqrt{\Delta~\Sigma~\Lambda~Q}
  (1+\sqrt{Q})\sin^2{\theta}} \Bigg{\{} \sin^2{\theta}(S_y \cos{\phi}-S_x \sin{\phi})
  \Delta^{3/2} \bigg{[}-\frac{\partial \mu}{\partial r} (\sqrt{Q}+1)\frac{P_\theta}{m} \nonumber \\
   &-& \frac{\partial \mu}{\partial \cos\theta}\frac{P_r}{m}\sin{\theta}+\sqrt{Q}
  \bigg{(}\frac{\partial \nu}{\partial r}\frac{P_\theta}{m}+\sin{\theta}
  (\frac{\partial \nu} {\partial \cos\theta}-\frac{\partial \mu} {\partial \cos\theta})
  \frac{P_r}{m} \bigg{)}  \bigg{]}  \nonumber \\
  &+& \frac{\Delta~\Sigma(2 \sqrt{Q}+1)\sin{\theta}~P_\phi}
  {m\sqrt{\Lambda}}\bigg{[}\sqrt{\Delta}~\frac{\partial \nu}{\partial r} \Big{(}-\cos{\theta}(S_x \cos{\phi}+S_y
  \sin{\phi})+S_z \sin{\theta}  \Big{)}  \nonumber \\
  &-& \frac{\partial \nu} {\partial \cos\theta}(S_x \sin\theta~\cos\phi+S_y \sin\theta~\sin\phi+S_z \cos\theta)\sin\theta \bigg{]} \nonumber \\
  &+& \Sigma \sqrt{\frac{\Delta}{\Lambda}}(r-M-\sqrt{\Delta})(\sqrt{Q}+1)\sin\theta\frac{P_\phi}{m}
  \bigg{[}\cos\theta(S_x\cos\phi+S_y\sin\phi)-S_z\sin\theta \bigg{]}
  \Bigg{\}}~~,
 \end{eqnarray}
  and the Hamiltonian providing the spin spin coupling reads  
 \end{widetext}

 \begin{widetext}
   \begin{eqnarray}\label{eq:RHamSSBL}
    H_{SS} &=&  \omega S_z+\sqrt{\frac{\Lambda}{\Delta}}\frac{\partial \omega}{\partial r}
    \frac{1}{2\Sigma^2\sqrt{Q}(1+\sqrt{Q})\sin^2\theta}\Bigg{\{}
    \frac{\Sigma~\Delta}{\sqrt{\Lambda}}\sin^2\theta(S_y\cos\phi-S_x\sin\phi)\frac{P_\phi P_\theta}{m^2} \nonumber \\
    &+& \frac{\Delta~\Sigma^2}{\Lambda}\sin\theta \left[-\cos\theta(S_x \cos\phi+S_y \sin\phi)
       +S_z\sin\theta\right]\frac{{P_\phi}^2}{m^2}  \nonumber \\
    &+& \Sigma~\Delta \sqrt{Q}(1+\sqrt{Q})\sin^3\theta\left[-\cos\theta(S_x \cos\phi+S_y \sin\phi)
       +S_z\sin\theta\right] \nonumber \\
    &+& \Delta^{3/2} \sin^3\theta \frac{P_r}{m^2}\Big{\{}\sqrt{\Delta}
  \big{[}\cos\theta(S_x\cos\phi+S_y\sin\phi)-S_z\sin\theta \big{]}P_r
  -(S_x \sin\theta~\cos\phi+S_y \sin\theta~\sin\phi+S_z \cos\theta)P_\theta\Big{\}}
  \Bigg{\}}  \nonumber \\
   &+& \frac{\sqrt{\Lambda}}{2\Sigma^2 \Delta \sqrt{Q}(1+\sqrt{Q})}\frac{\partial\omega}{\partial \cos\theta}
  \Bigg{\{}
  -\frac{\Delta~\Sigma^2}{\Lambda}\frac{{P_\phi}^2}{m^2}(S_x \sin\theta~\cos\phi+S_y \sin\theta~\sin\phi+S_z \cos\theta) \nonumber \\
   &+& \frac{\Sigma~\Delta^{3/2}}{\sqrt{\Lambda}}\frac{P_r P_\phi}{m^2}\sin\theta
   (S_y\cos\phi-S_x\sin\phi)
   + \sin^2\theta \Delta\Big{\{}(S_x \sin\theta~\cos\phi+S_y \sin\theta~\sin\phi+S_z \cos\theta)\nonumber \\
   &\times&\left(\frac{P_\theta^2}{m^2}-\Sigma\sqrt{Q}(1+\sqrt{Q})\right)
   + \sqrt{\Delta}\frac{P_\theta P_r}{m^2} [-\cos\theta(S_x\cos\phi+S_y\sin\phi)+S_z\sin\theta] \Big{\}}
  \Bigg{\}}~~,
  \end{eqnarray}
 \end{widetext}
 where the $S_I$ is written in the corresponding cartesian coordinates, i.e.,
 \begin{eqnarray} \label{eq:CtoBL}
  x &=& r \sin{\theta} \cos{\phi}~~, \nonumber  \\
  y &=& r \sin{\theta} \sin{\phi}~~, \nonumber  \\
  z &=& r \cos{\theta} ~~,
 \end{eqnarray}
 and $\omega,~\mu,~\nu,~Q$ are the following functions
 \begin{eqnarray}
  \omega &=& \frac{2 a M r}{\Lambda}~~, \nonumber  \\
  e^{2\nu} &=& \frac{\Delta\Sigma}{\Lambda}~~, \nonumber \\
  e^{2\mu} &=& \frac{4 \Sigma}{(r-M+\sqrt{\Delta})^2}~~,  \nonumber \\
  Q &=& 1+\frac{\gamma^{ij}}{m^2} P_i P_j~~.
 \end{eqnarray}
 For more about the canonical Hamiltonian formalism and the derivation of the above
 Hamiltonian function see \cite{Barausse09} and \cite{Barausse10} respectively.
 
 The equations of motion for the canonical variables as a function of the coordinate
 time $t$ read
 \begin{align}
  \frac{d x^i}{dt} &=\pd H{P_i}~~,\nonumber \\
  \frac{d P_i}{dt} &=-\pd H{x^i}~~,\nonumber\\
 \frac{d S_I}{dt} &=\epsilon_{IJC}\pd H{S_J}S^C~~\label{eq:EqMHam}~~,
 \end{align} 
 where $\epsilon_{IJL}$ is the Levi-Civita symbol.
 
\section{2D and 4D  Poincar\'{e} sections} \label{sec:2D4DPs} 
 
 \subsection{The issue of integrability} \label{sec:IntCh}
 
 The canonical Hamiltonian approximation provided in \cite{Barausse09} has five
 degrees of freedom. Three degrees of freedom come from the coordinates,
 and two degrees from the spin vector \cite{KLLS}. In \cite{KLLS} it has been
 shown that for the Schwarzschild spacetime background the Hamiltonian
 approximation possesses five integrals of motion. The spherically symmetric
 background corresponds to the preservation of the total angular momentum, thus,
 two independent and in involution integrals come from the spherical symmetry; 
 since the Hamiltonian is autonomous, the Hamiltonian function is a constant of
 motion, representing the energy; the measure of the particle's spin is conserved,
 and the measure of the orbital angular momentum is a constant as well. Hence, 
 since we have five independent and in involution integrals for five degrees
 of freedom, the Hamiltonian of a spinning particle for a Schwarzschild 
 background is integrable \cite{KLLS}.
 
 For nonzero spin of the central black hole, however, chaotic motion appears
 (see figure~3 of \cite{KLLS}). This means that for the Kerr background the
 revised Hamiltonian of \cite{Barausse10} is non-integrable. Actually, figure~3
 of \cite{KLLS} is a projection of a 4D Poincar\'{e} map on a 2D surface of
 section. The 2D projections of a 4D Poincar\'{e} map is an old technique 
 to visualize the dynamics of a chaotic system (method \ref{item:2Dpr} in 
 appendix~\ref{sec:V4D}). Similar techniques have been employed in previous 
 studies \cite{Suzuki97,Hartl03a,Hartl03b,Han08} when the question of chaos 
 was examined for spinning particles using MP equations. However, since the 
 MP equations are not symplectic, the use of surface of sections for studying 
 their dynamics is ambiguous. On the other hand, the canonical Hamiltonian 
 formalism of \cite{Barausse09} is symplectic (see, e.g., appendix A in 
 \cite{KLLS}), and, hence, the subsequent study of Poincar\'{e} sections 
 stands on solid ground from this point of view.
 
 \subsection{Setting up the numerics} \label{sec:SetNum}
 
 To evolve the Hamiltonian equation of motion \eqref{eq:EqMHam} we need to 
 set up  the initial conditions of our system. We have nine variables, i.e., 
 three variables for the position, three for the momentum, and three for the 
 spin. In the case of the Kerr background we have two integrals of motion 
 apart from the Hamiltonian function $H$ \eqref{eq:HamSP}. Namely, the 
 azimuthal component of the total angular momentum \cite{Barausse10,KLLS}
 \begin{equation} \label{eq:Jz}
   J_z=P_\phi+S_z~~,
 \end{equation}
 and the measure of the particle's spin \cite{Barausse09}
 \begin{equation} \label{eq:SpinM}
   S=\sqrt{S_x^2+S_y^2+S_z^2}~~,
 \end{equation}
 are preserved. For a group of orbits to belong to the same surface of section 
 they have to share the same values of $J_z$, $S$ and $H$. Thus, we are going to
 use the above three constants to fix the initial conditions. 
 
 Since the Kerr background is axisymmetric, the initial value of the azimuthal 
 angle $\phi$ can be set to $0$ without loss of generality. The equatorial
 plane $\theta=\pi/2$ seems to define the appropriate surface of section,
 due to reflection symmetry along the equatorial plane of the Kerr spacetime.
 The equatorial plane was also chosen as the surface of section by previous studies
 of the spinning particle dynamics \cite{Suzuki97,Hartl03a,Hartl03b,Han08}.
 On the equatorial plane we choose initial conditions along the
 radial direction $r$ and for each orbit we set the initial radial momentum to
 $P_r=0$. The spin components $S_x,~S_y$ are chosen to be set to $0$, and, thus,
 the measure of the component $S_z$ is defined by the spin's magnitude. The sign
 of $S_z$ shows if the particle's spin is initially aligned with the spin of the
 central object (positive sign) or anti-aligned (negative sign).
 From \eqref{eq:Jz}, with given $S_z$, we can get $P_\phi$, while $P_\theta$ is
 found through a Newton iteration for a given value of the Hamiltonian function $H$
 \footnote{With all the other phase space variables fixed as explained in the text, 
 the Hamiltonian function can be rewritten as an effective function of $P_\theta$
 alone which drastically reduces the complexity of the Newton iteration.}.
 Obviously the above initial condition setting is not unique, but we found it
 convenient for our investigation.
 
 The equations of motion~\eqref{eq:EqMHam} are evolved by a Gauss Runge--Kutta 
 integration scheme which has very good conservation properties for symplectic
 systems (see, e.g.,~Appendix A in~\cite{LGSK}). On the surface of section we 
 record crossings with $P_\theta>0$. In order to calculate the phase space 
 points on the sections very precisely, we take use of the integration scheme's
 interpolation property as described in Appendix A of~\cite{KLLS}.
 
 In our visualization we are going to use only the variables 
 $r,~P_r,~P_\theta~P_\phi$, since by using the constants of
 motion~\eqref{eq:Jz}-\eqref{eq:SpinM} we can reduce our phase space to the
 positions, and the momenta. Above we have chosen $\theta=\pi/2$
 for our surface of section due to the reflection symmetry. Moreover, even if
 $\phi$ evolves in time, we do not to use it for the $4D$ Poincar\'{e} sections,
 because the Kerr spacetime is axially symmetric and, therefore, the variable
 $\phi$ should not carry any useful information. Thus, in our 4D~Poincar\'{e}
 sections we are using $r,~P_r,~P_\theta$ for the 3D projection, while $P_\phi$
 is represented by the color. However, note that due to the
 constant~$J_z$~\eqref{eq:Jz}, the use of $P_\phi$ to color the consequents is
 equivalent to the use of $S_z$, i.e., the maxima of the one quantity correspond
 to the minima of the other one. 
 
 The spin is measured in $m~M$ units, namely $S/(m M)$ is dimensionless. By
 setting $m=M=1$ the spin is dimensionless, and all the other quantities as well.
 In some of our numerical examples we are using unrealistic high values for the
 particle's spin measure, e.g. $S\approx 1$. However, these values are
 dynamically valid even for the linearized in spin Hamiltonian
 formulation we are using, because once the Hamiltonian function is explicitly
 written the Hamiltonian system is selfconsistent. Namely, the Hamiltonian
 function itself depends just linearly on the spin components, and the Hamiltonian
 equations~\eqref{eq:EqMHam} are just linearly depended on the spin as well.
 The only limitation is the astrophysical. The dimensionless spin value becomes
 astrophysically relevant for EMRI when $S<10^{-4}$ (for more details see 
 section II.B in \cite{Hartl03a}). However, one has to keep in mind that the
 aim of this work is basically a dynamical investigation of the system, not an
 astrophysical one.  
 
 As far as the Kerr parameter is concerned, we have chosen the value $a=0.9$ in
 our study. The reason is the following. In order to have integrability
 we can go either to the geodesic limit $(S=0)$, or to the Schwarzschild limit
 $(a=0)$. Thus, in order to have the most pronounced non-integrability effects,
 we have to go away from both above limits, which is the case with $a=0.9$. This
 does not mean that for smaller Kerr parameters we cannot find signs of chaos.
 Actually, the non-integrability of the linear in spin Hamiltonian approximation
 for the Kerr background was found for $a=0.1$ (figure~3 in \cite{KLLS}).
  
  \begin{figure}[htp]
  \centerline{ \includegraphics[width=0.45\textwidth]{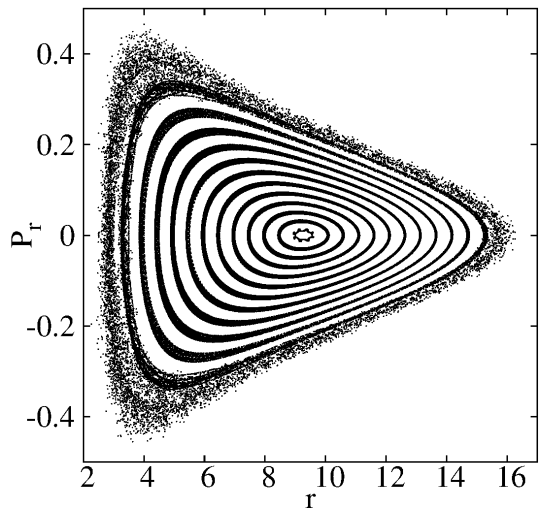}}
  \caption{A 2D projection of a Poincar\'{e} section on the $r,~P_r$ plane for 
  spins $a=0.9,~S=1$ and $H=0.95,~J_z=2$.
  }
  \label{Fig:2DS1}
 \end{figure}
 
 \subsection{Examples for $S=1$} \label{sec:ExS1}
 
 In order to find chaos we use the extreme case of $S=1$ in our first example.
 Fig.~\ref{Fig:2DS1} shows a two dimensional projection of a Poincar\'{e}
 section. We can observe a chaotic region (scattered points)
 encircling an island of stability. The chaotic sea is confined between two
 surfaces. The inner one, which defines the limit of the island of stability, is a
 KAM torus, while the outer one is the boundary of the allowed motion. The outer
 boundary is indicated by the outer limit of the chaotic orbit.
 The boundary of the allowed motion has an opening around $r=2,~P_r=0$ from
 which the chaotic orbits are plunging towards the central black hole $(r=0)$.
 However, our observations are not unambiguous, since we do not see a Poincar\'{e}
 section in Fig.~\ref{Fig:2DS1}, but a projection. A 2D Poincar\'{e} section is
 accurate only for a Hamiltonian system of two degrees of freedom. In a two
 degrees of freedom system the KAM curves have zero width, and chaotic regions
 are represented by scattered dots covering a nonzero width region.
 In Fig.~\ref{Fig:2DS1} we see KAM tori projected on a 2D plane, so the width of
 the KAMs is nonzero. Thus, a 2D Poincar\'{e} projection does not offer an
 unambiguous criterion to distinguish chaos from order. In order to drive safe
 conclusions we have to use 4D Poincar\'{e} sections.

 \begin{figure}[htp]
  \centerline{ \includegraphics[width=0.49\textwidth]{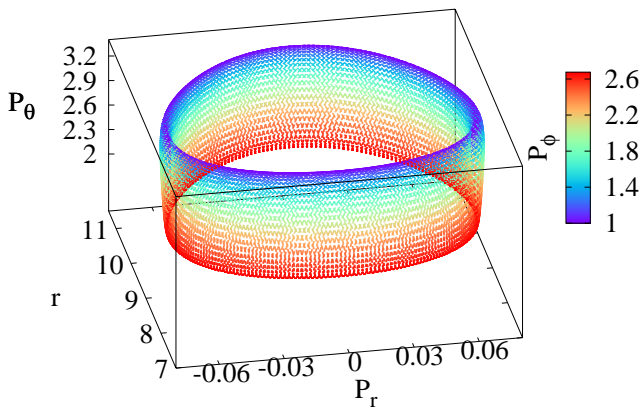}}
  \caption{ A regular torus from Fig.~\ref{Fig:2DS1} with initial $r=7.5$ 
  on a 4D Poincar\'{e} section.  
  }
  \label{Fig:4DS1reg}
 \end{figure}
 
 Using the technique of color and rotation on a 4D Poincar\'{e} section, the
 regularity of an orbit in the neighborhood of a stable periodic orbit is
 indicated in the topology of the 3-dimensional projection by the presence of a
 torus, with a smooth color variation on its surface. This is determined by the
 distribution of the consequents in the 4th dimension \cite{Patsis94}. We use
 the orbit starting from $r=7.5$ in Fig.~\ref{Fig:2DS1} to give our first
 example of a regular orbit on a 4D Poincar\'{e} section (Fig.~\ref{Fig:4DS1reg}). 
 In Fig.~\ref{Fig:4DS1reg} we observe that as the orbit evolves on the
 rotational torus projected on the $r,~P_r,~P_\theta$ surface, the colors
 representing $P_\phi$ vary smoothly \cite{Katsanikas11a}.
 This means that the orbit is regular. We use the software package ``gnuplot''
 to visualize our results. We give the viewing angles of the 3D projections for 
 Fig.~\ref{Fig:4DS1reg} and all the subsequent similar figures of our paper in
 Table~\ref{tab:ViewAngles}.
 
  \begin{figure*}[htp]
  \centerline{ \includegraphics[width=0.49\textwidth]{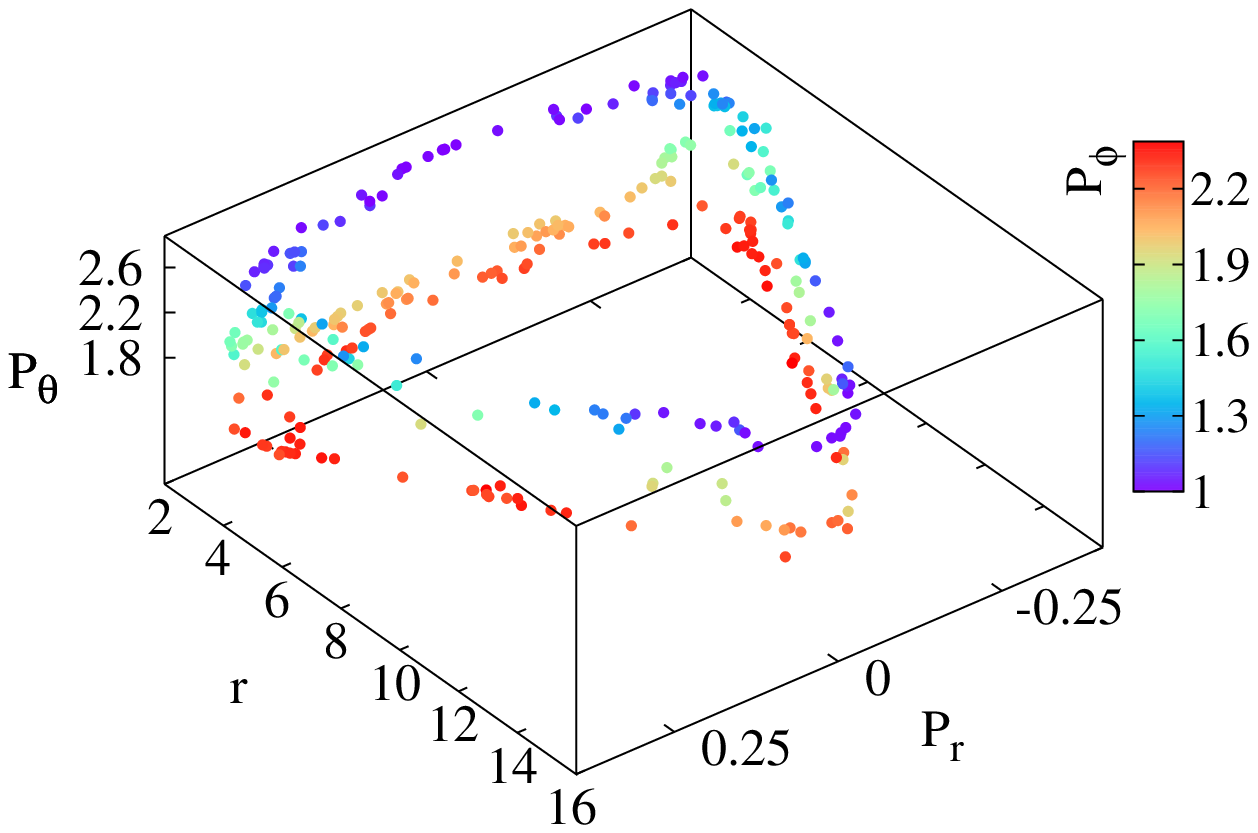}
  \includegraphics[width=0.49\textwidth]{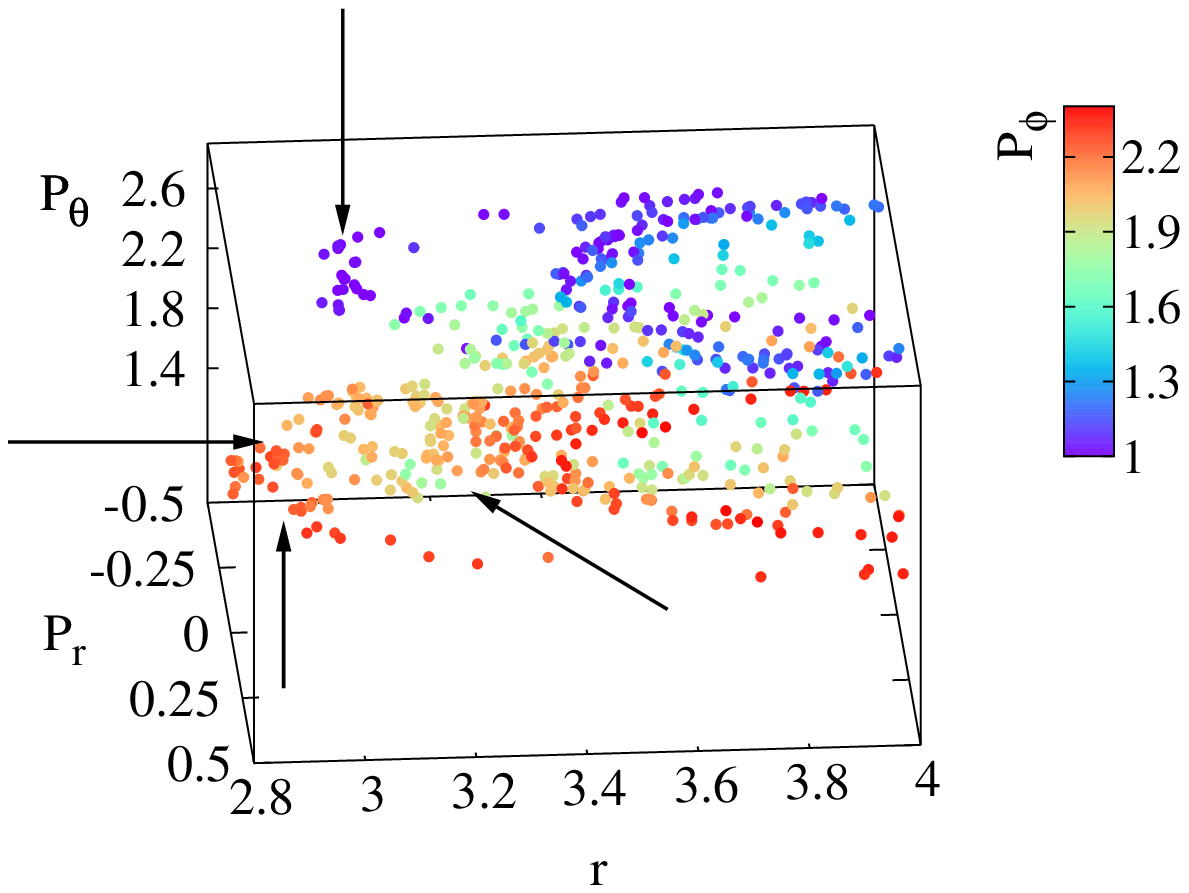}}
  \caption{ A chaotic orbit with initial $r=3,~P_r=0$ in Fig.~\ref{Fig:2DS1}
  depicted in a 4D Poincar\'{e} section. The left plot shows the initial $300$ 
  crossings of the orbit through the Poincar\'{e} section, while the right shows
  a detail from the Poincar\'{e} section when $1500$ crossings have been reached. 
  The arrows indicate consequents almost on the surface that separates the
  allowed from the non-allowed space for the motion of the particle. For further
  explanations see text. 
  }
  \label{Fig:4DS1ch}
 \end{figure*}

 On a 4D Poincar\'{e} section the chaotic nature of an orbit is demonstrated by
 its irregular behavior on the 3-dimensional projection or/and by the mixing of 
 colors representing the 4th dimension. In Fig.~\ref{Fig:4DS1ch} we consider the
 chaotic orbit starting from $r=3$ on the 2D projection in Fig.~\ref{Fig:2DS1}.
 Initially the orbit sticks around a KAM lying on the border of the island of
 stability (This is given in  the left plot of Fig.~\ref{Fig:4DS1ch}). By
 sticking around the torus it mimics a regular orbit (the color variation is
 smooth), but as the orbit evolves it departs from the KAM torus and sticks on
 the surface that defines the space for the allowed motion. The consequents of
 the orbit exhibit a smooth color variation. This is typical of the phenomenon
 of stickiness and it is quite common for weakly chaotic orbits, which are
 called sticky, see, e.g., \cite{Katsanikas11a}. The arrows at the right plot
 in Fig.~\ref{Fig:4DS1ch} show points that stick in this case on the outer 
 boundary. The chaotic nature of the orbit is defined by its irregular behavior,
 and not by the color mixing. The orbit, after 1500 consequents, does not form a
 torus with small color variation on it like in Fig.~\ref{Fig:4DS1reg}, but it
 has a  double loop structure. The fact that we do not have color mixing
 indicates stickiness \cite{Katsanikas13}. This behavior is similar to a
 weakly chaotic orbit that is trapped between two invariant curves in the case
 of a 2D Hamiltonian System.
 
 \subsection{Examples for $S=\sqrt{0.1}$} \label{sec:ExSroot0p1}
 
  \begin{figure}[htp]
  \centerline{ \includegraphics[width=0.45\textwidth]{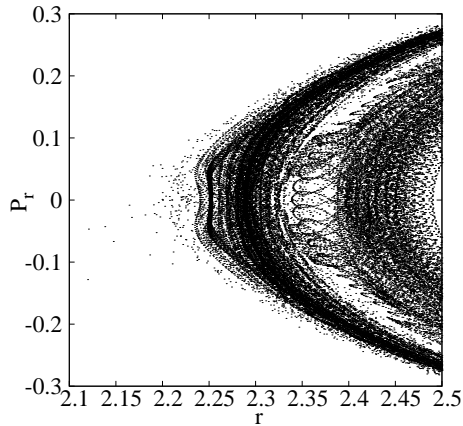}}
  \caption{A detail from a 2D projection of a Poincar\'{e} section on the
  $r,~P_r$ plane for spins $a=0.9,~S=\sqrt{0.1}$ and $H=0.95,~J_z=2$.}
  \label{Fig:2DSroot0p1}
 \end{figure} 
 
 We keep the same energy and angular momentum as in Sec.~\ref{sec:ExS1}, but
 we reduce the spin measure to $S=\sqrt{0.1}$. In this case a 2D projection of
 the whole phase space like the one in Fig.~\ref{Fig:2DS1} is hardly discernible
 from a proper Poincar\'{e} map coming from a system with 2 degrees of freedom.
 One has to focus on a small region of the phase space to see the real structure
 (Fig.~\ref{Fig:2DSroot0p1}). In Fig.~\ref{Fig:2DSroot0p1} we observe that there
 is still a chaotic region surrounding the main island of stability (scattered
 points on the left side of the plot), and that the KAM tori have nonzero width.
 It is worth mentioning that in a system of 3 degrees of freedom the chaotic
 regions communicate even if we see KAMs between them in the 3D projections of
 the 4D space of section. On the contrary in  2~degrees of freedom systems,
 when a KAM is lying between two chaotic regions in the 2D surface of section
 it does not allow them to communicate\footnote{By ``communicate'' we mean that
 a chaotic orbit can go from the one region to the other.}. A case where the two
 chaotic regions communicate is given in Fig.~\ref{Fig:2DSroot0p1}. Apart from
 the outer chaotic region there is a chaotic region lying on the interval 
 $2.35 \lesssim r \lesssim 2.37$. This region is inside  the KAM  tori that are 
 lying on the interval $2.24 \lesssim r \lesssim 2.25$ on the $P_r=0$ line in 
 Fig.~\ref{Fig:2DSroot0p1}. By starting integrating an orbit in the inner
 chaotic region we soon end up in the outer one, since the two regions
 communicate.
 
 \begin{figure}[htp]
  \centerline{ \includegraphics[width=0.45\textwidth]{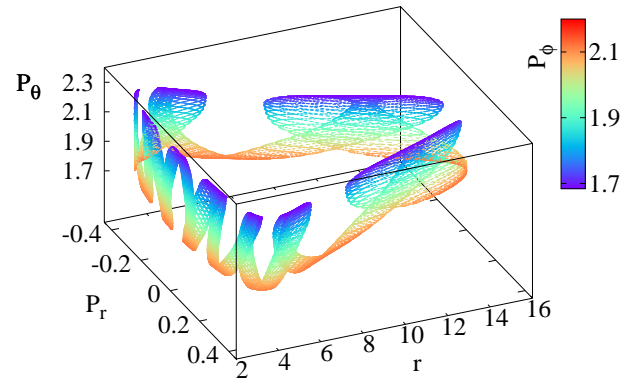}}
  \caption{A regular torus from Fig.~\ref{Fig:2DSroot0p1} with initial $r=2.4$ 
  on a 4D Poincar\'{e} section.  
  }
  \label{Fig:4DSroot0p1reg}
 \end{figure}
 
   \begin{figure}[htp]
  \centerline{ \includegraphics[width=0.45\textwidth]{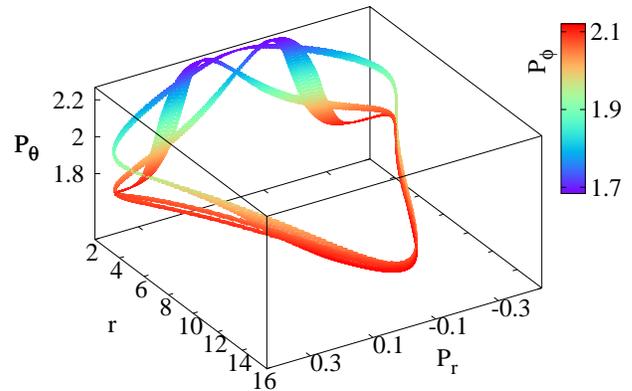}}
  \caption{Filament corresponding to a chaotic orbit from
  Fig.~\ref{Fig:2DSroot0p1} with initial  $r=2.32$ on a 4D Poincar\'{e} section. 
  The consequents after long integration time diffuse i phase space.
  }
  \label{Fig:4DSroot0p1fil}
 \end{figure}
 
 Actually the structure of the phase space is far more complicated. An example
 is a regular orbit, starting from  $r=2.4$  which is represented by a structure
 that looks like nooses in a row in the 2D subspace $(r,P_r)$ of the 
 4D~Poincar\'{e} section (Fig.~\ref{Fig:2DSroot0p1}). This regular orbit is 
 represented by a warped rotational torus on the 4D~Poincar\'{e} section 
 (see, e.g., \cite{Vrahatis97}, \cite{Katsanikas11a}). In 
 Fig.~\ref{Fig:4DSroot0p1reg} we see the real structure of the warped  
 rotational torus. The regular orbit follows the  warping of the torus while the
 color varies smoothly during the time of integration. On the other hand, weakly
 chaotic orbits lie in the region which is apparently dominated by KAM tori
 ($2.24 \lesssim r \lesssim 2.25$ in Fig.~\ref{Fig:2DSroot0p1}).
 In Fig.~\ref{Fig:4DSroot0p1fil} we plot such a weakly  chaotic orbit. It is
 represented by a 3D filamentary structure with self-intersections in the 3D 
 subspace $(r,P_r,P_{\theta})$ of the 4D space of section. We observe that this
 structure has smooth color variation and that we have the same color (the same
 value in the 4th dimension) at the regions of the self-intersections.
 
 We underline the fact that in Fig.~\ref{Fig:4DSroot0p1fil} we observe two 
 self-intersections that do not have the same color. If we rotate the figure, 
 we can see these self-intersections from different viewing angles and we can 
 observe very easily that these self-intersections do not exist in the 3D subspace.
 This means that these self-intersections are due to the viewing angles and they
 do not really exist. The smooth color variation of the 3D~filamentary shows
 that the 4th dimension supports the geometry of this structure in the 4D space 
 of section. This also gives us the dynamical information that these
 self-intersections occur in the 4D space. Such 4D~filamentary structures have
 been encountered for the first time in a 3D~galactic Hamiltonian system 
 in \cite{Katsanikas11c} and they are found at the neighborhood of unstable
 periodic orbits with high multiplicity \cite{Katsanikas11c}. The orbits that 
 are represented by these structures are sticky chaotic orbits.
 
 Such weakly chaotic orbits have as a 2D counterpart the chaotic orbits that can
 be found in chains of elliptic and hyperbolic points in resonance zones. These
 chaotic orbits connect the hyperbolic points and surround
 the islands of stability of the elliptic ones.
 In the case we study here, these weakly chaotic orbits extend into the 3D space 
 of the projection, while they have a smooth color variation along the filament 
 they form. However, if we continue the integration for long times, the orbit
 will diffuse in the 4D space, something that will be demonstrated clearly in 
 the next example.

 \subsection{Examples for $S\le 0.1$} \label{sec:ExSle0p1}
 
 \begin{figure}[htp]
  \centerline{ \includegraphics[width=0.45\textwidth]{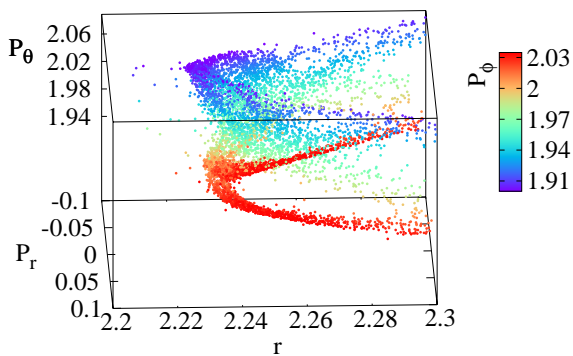}}
  \caption{A detail from a 4D Poincar\'{e} section of a filamentary chaotic
  orbit with $S=0.1$ starting from $r=2.225$.  
  }
  \label{Fig:4DS0p1fil}
 \end{figure}
 
 If we reduce the spin measure to $S=10^{-1}$ and $S=10^{-2}$, we encounter 
 again 4D tori and 4D filamentary structures in the 4D space of section that
 correspond to regular orbits and sticky chaotic orbits respectively. In 
 Fig.~\ref{Fig:4DS0p1fil} (for $S=0.1$) we observe a 4D filamentary structure. 
 Despite the fact that we have smooth color variation for the 4th dimension
 $P_{\phi}$ the consequents depart from this filamentary structure through the
 3D subspace $(r,P_r,P_{\theta})$, and they occupy larger volumes in the phase 
 space (before the final plunge towards the black hole). These consequents can 
 be observed at the left side of Fig.~\ref{Fig:4DS0p1fil}. The departure of
 these points from the filamentary structure happens earlier than in the case
 described in Fig.~\ref{Fig:4DSroot0p1fil}. However, in both cases we observe
 stickiness on 4D Poincr\'{e} sections in structures that correspond to chaotic
 zones around unstable periodic orbits with high multiplicity for the first time
 in a relativistic system.
   
  \begin{figure}[htp]
  \centerline{ \includegraphics[width=0.45\textwidth]{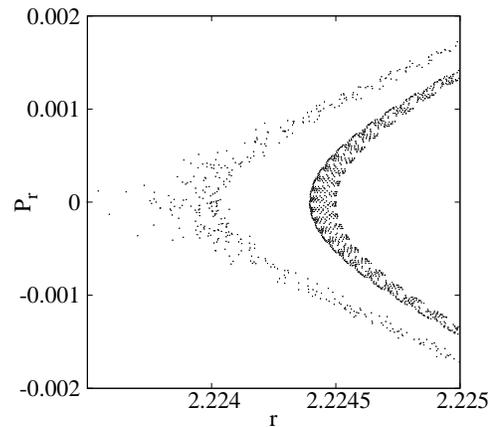}}
  \caption{A detail from a 2D projection of a Poincar\'{e} section on the 
  $r,~P_r$ plane for spins $a=0.9,~S=0.001$ and $H=0.95,~J_z=2$.  
  }
  \label{Fig:2DS0p001}
 \end{figure}
 
 The last significant imprints of chaos are found for $S=10^{-3}$. For such
 low value of spin the 2D projection of a Poincar\'{e} section shown in
 Fig.~\ref{Fig:2DS0p001} is very close to what one would expect to see in a case
 of a system with 2 degrees of freedom. In Fig.~\ref{Fig:2DS0p001}, we see a
 chaotic zone (scattered points on the left side), and a KAM torus 
 (orbit on the right side of the plot). We have to focus significantly on the 
 surface of section in order to make apparent the chaotic zone and the width of
 the torus. 
 
 For spins $S\leq 10^{-4}$ the presence of chaos appears to be negligible, and
 if this is the case it can be practically ignored. Even non-integrability effects
 like the existence of islands of stability near resonances can be neglected for
 any practical reason as well. In few words the system is nearly integrable, in
 agreement with the recent findings of \cite{Ruangsri15}, where no traces of
 resonant orbits were found in a study of a the linearized in spin MP equations.
 It is worth reminding that $S\leq 10^{-4}$ is the upper limit for the EMRIs, and
 it is interesting to notice that this value is also the upper limit for which
 the orbits produced by the Hamiltonian approximation start to match the orbits
 produced by the MP equations with NW SSC \cite{LGSK}.
 
 \begin{table}

  \begin{tabular}[t]{c | c c }
  Fig. & $\theta$ & $\phi$  \\ 
  \hline
  \ref{Fig:4DS1reg} &  $47^{o}$ &$349^{o}$  \\
  \ref{Fig:4DS1ch}~(left panel) &  $36^{o}$ &$144^{o}$  \\
  \ref{Fig:4DS1ch}~(right panel)&  $136^{o}$& $84^{o}$ \\
  \ref{Fig:4DSroot0p1reg} &  $46^{o}$ & $66^{o}$ \\
  \ref{Fig:4DSroot0p1fil} &  $44^{o}$ & $148^{o}$ \\
  \ref{Fig:4DS0p1fil} & $60^{o}$ & $88^{o}$  \\
  \end{tabular} 
  \caption{The view points of the figures, which are depicting 4D~Poincar\'{e}
  section, are given in spherical coordinates $(\theta,\phi)$ as defined in the
  gnuplot software package.}
  \label{tab:ViewAngles}
 \end{table}

 \section{Discussion and conclusions} \label{sec:ConDis}

 The method of color and rotation \cite{Patsis94} is used for the first time in
 a relativistic system. Until now this method was used in 3D galactic
 Hamiltonian systems 
 (\cite{Katsanikas11a,Katsanikas11b,Katsanikas11c,Patsis14a,Patsis14b},
 the 3D circular restricted three body problem \cite{Geisel13} and 
 a 4D symplectic map \cite{Zachilas13}. We encountered three types of orbits
 in our study, which, though studied in detail in a 3D galactic system
 \cite{Katsanikas11a,Katsanikas11c}, have never been investigated in other 3D
 systems in the framework of general relativity. These three types of orbits are:

 \begin{enumerate}

  \item The first type of orbits are the regular orbits. These orbits are 
  represented  on the 4D Poincar\'{e} spaces of section by 4D rotational tori 
  \cite{Katsanikas11a,Vrahatis97}. These tori have the topology of a regular 
  torus in the 3D projections of the 4D Poincar\'{e}  space of section. Some of
  them are smooth regular tori and few of them are warped. Nevertheless, all of 
  them manifest smooth color variation on them.

  \item The second type of orbits are chaotic orbits that initially stick on 4D
  rotational tori (on the 4D Poincar\'{e} section), before they diffuse in the
  phase space.

  \item The third type of orbits are a special case of chaotic orbits. They are
  represented by 4D filamentary structures on the 4D Poincar\'{e} sections as in 
  \cite{Katsanikas11c}. These structures are in the neighborhood of unstable
  periodic orbits with high multiplicity. Such orbits are sticky chaotic orbits
  since their consequents leave the 4D filamentary structures after a longer
  time of integration. 

 \end{enumerate} 

 In general we did not encounter strong chaos in the system, which would be
 manifested by color mixing on the 4D Poincar\'{e} sections. We encountered only
 weakly chaotic and sticky orbits. Moreover, we observe that
 chaotic motion seems to be insignificant, and its contribution to the overall
 dynamics can be probably be neglected, when the dimensionless spin becomes
 smaller than $S= 10^{-4}$, i.e. when the value of the spin is in the
 astrophysical relevant interval for extreme mass ratio inspirals.
 However, from a dynamical point of view the inclusion of the particle's spin
 in the motion of a small compact object is just one way to go from the
 integrable case of geodesic motion on a Kerr black hole background to a
 non-integrable system. For example, it is well known that rings and halos
 around black holes can induce chaotic motion (see, e.g., \cite{SemSuk}). The
 same effect takes place when the spacetime around the central supermassive
 object is described by a non-Kerr black hole (see, e.g., \cite{nonKerr}).  
 Non-integrability can also originate from the self-force or from the inclusion
 of the quadrupole momentum to the Mathisson-Papapetrou equations. In few words,
 there are many reasons for a extreme mass ratio binary to be described by a
 non-integrable system. However, it is unclear to which extent the effects
 coming from the non-integrability can affect the motion of the small body.

\begin{acknowledgments}
G.L-G is supported by UNCE-204020a and by GACR-14-10625S. This work was partially
supported by Research Committee of the Academy of Athens (project 200/854).
We would like to thank Prof. GeorgevContopoulos for carefully reading the
manuscript and for his useful suggestions.

\end{acknowledgments}

\appendix

 \section{Visualizing 4D Poincar\'{e} sections} \label{sec:V4D}

 Several methods have been used for visualizing the 4D spaces of section:
  
 \begin{enumerate}
  \item \label{item:2Dpr}
   {\bf 2D and 3D projections}: In this method the points of an orbit are
   plotted in a 2D subspace \cite{Contopoulos89,Skokos97}, \S 2.11.11 in
   \cite{Contopoulos02} or in a 3D subspace \cite{Vrahatis96,Vrahatis97} of the 
   4D Poincar\'{e} space of section. This method has the disadvantage that the
   distribution of the points in the 4 dimensional space is lost. However, in 
   many cases, thin structures resembling invariant curves in the 2D case
   indicate the presence of tori. 

  \item {\bf Stereoscopic Views}: Stereoscopic views of a 3D subspace of the
   4D Poincar\'{e} space of section are used in order to understand the topology
   of the 3D projections \cite{Froeschle70,Martinet81,Contopoulos82} of the
   figures. For this reason, two figures are needed, one for each eye of the
   observer. However, this method cannot give any information about the behavior
   of the orbit in the 4th dimension.   

  \item {\bf The Method of Slices}: In this method \cite{Froeschle70,Froeschle72}
   2D slices for different values of the third dimension of a 4D Poincar\'{e}
   space of section are produced. The successive 2D figures help one to see the
   distribution of the points of an orbit in the 3D subspace of the 4D space of
   section. By using this method many figures are needed in order to understand 
   the third dimension and the fourth dimension in the 4D space of section is
   absolutely lost. An improved version of this method has been demonstrated 
   recently in \cite{Richter14,Lange14}. In this case 3D are used, instead of 2D,
   slices of the space of section and they can be rotated by using standard 3D
   graphics software. By doing so, one can better see the third dimension and
   ``visualize'' the fourth dimension of the 4D Poincar\'{e} space of section as
   well. The disadvantage of this version of the slices method is that the 3D
   slices can be very complicated and sometimes it is difficult to see directly
   the topology of the orbits in the 4D Poincar\'{e} space of section.      

  \item {\bf The Method of Color and Rotation}: This method was introduced in 
  \cite{Patsis94} and is applied in the present paper (see also our
  introduction). The method has the advantage that we can observe the 4D 
  distribution of the points of an orbit without any change of the 3D geometry
  or the 3D topology of the orbit in the 4D Poincar\'{e} space of section. 

\end{enumerate}


\begin{thebibliography}{9}

\bibitem{LISA} 
P., Amaro-Seoane, S. Aoudia, S. Babak,P. Bin\'{e}truy, E. Berti, A. Boh\'{e},
C. Caprini, M. Colpi, N.~J. Cornish, K. Danzmann, J.-F. Dufaux, J. Gair, 
I. Hinder, O. Jennrich, P. Jetzer, A. Klein, R.~N Lang, A. Lobo, T. Littenberg, 
S.~T. McWilliams, G. Nelemans, A. Petiteau, E.~K Porter, B.~F. Schutz,
A. Sesana, R. Stebbins, T. Sumner, M. Vallisneri, S.  Vitale, M. Volonteri,
H. Ward, B. Wardell, {\it GW Notes} {\bf 6}, 4-110 (2013)

 \bibitem{Mathisson37}
 M. Mathisson, {\it Acta Phys. Polonica} {\bf 6}, 163 (1937)
%
 \bibitem{Papapetrou51}
 A. Papapetrou, {\it Proc. R. Soc. London  Ser. A} {\bf 209},
 248 (1951)
%
\bibitem{Moeller49} C. M\o{}ller,\textit{ Annales de l'I. H. P.
}\textbf{11}, 251 (1949)
%
 \bibitem{Pirani56} F. A. E. Pirani, \textit{Acta Phys. Polonica}
 \textbf{15}, 389 (1956)
%
 \bibitem{Tulczyjew59} W. Tulczyjew, \textit{Acta Phys. Polonica}
\textbf{18}, 393 (1959)
%
 \bibitem{Semerak99} O. Semer\'{a}k, \textit{Mon. Not. R. Astron. S.}
 \textbf{308}, 863 (1999)
%
 \bibitem{Kyrian07} K. Kyrian, and O. Semer\'{a}k, \textit{Mon. Not. R. Astron. S.}
 \textbf{382}, 1922 (2007)
%
\bibitem{Semerak15} 
 O. Semer{\'a}k,  and M. \v{S}r\'{a}mek, \textit{Phys. Rev. D} \textbf{ 92}, 
 064032 (2015)
 
%
 \bibitem{Costa12} L. F. Costa, C. Herdeiro, J. Natario and M. Zilh\~{a}o,
 \textit{Phys. Rev. D} \textbf{85}, 024001 (2012).
%
 \bibitem{Suzuki97} S. Suzuki and K. Maeda, \textit{Phys. Rev. D}
 \textbf{55}, 4848 (1997)
%
 \bibitem{Hartl03a} M.~D. Hartl, \textit{Phys. Rev. D} \textbf{67},
 024005 (2003)
%
 \bibitem{Hartl03b} M.~D. Hartl, \textit{Phys. Rev. D} \textbf{67},
 104023 (2003)
%
 \bibitem{Han08} W.~Han, \textit{Gen. Rel. Grav.} {\bf 40}, 1831 (2008)
%
 \bibitem{Rudiger} R.~R\"{u}diger, \textit{Proc. R. Soc. London Ser. A}
 \textbf{375}, 185 (1981); \textbf{385}, 229 (1982)
%
 \bibitem{Hinderer13} T. Hinderer, A. Buonanno, A. H. Mrou\'{e} et. al,
 \textit{Phys. Rev. D} \textbf{88},084005 (2013)
%
 \bibitem{Barausse09}
 E. Barausse, E. Racine, and A. Buonanno, {\it Phys. Rev. D} {\bf 80}, 104025 (2009)
%
 \bibitem{NewtonWigner49} T.~D. Newton, and E.~P. Wigner,
 \textit{Rev. Mod. Phys.} \textbf{21}, 400 (1949)
%
\bibitem{Barausse10}
 E. Barausse, and A. Buonanno,  {\it Phys. Rev. D} {\bf 81}, 084024 (2010) 
%
 \bibitem{KLLS} D. Kunst, T. Ledvinka, G. Lukes-Gerakopoulos, and J. Seyrich,
 \textit{Phys. Rev. D} \textbf{93}, 044004 (2016)
%
 \bibitem{Contopoulos90}
 G. Contopoulos, {\it Proc. Math. Phys. Sc.} {\bf 431}, 183 (1990) 
%
 \bibitem{Contopoulos89}
 G. Contopoulos, and B. Barbanis   {\it Celest. Mech. Dyn. Astron.} 
 {\bf 59}, 279-300 (1989)
%
 \bibitem{Vrahatis97}
  M.N. Vrahatis, H. Isliker, and T.C. Bountis {\it Int. J. Bif. Chaos} 
  {\bf 7}, 2707-2722 (1997)
%
 \bibitem{Froeschle70}
  C. Froeschl\'{e}   {\it Astron. Astrophys.} {\bf 4}, 115-128 (1970)
%
 \bibitem{Martinet81}
  L. Martinet,  and P. Magnenat  [1981]  {\it Astron. Astrophys.} {\bf 96}, 
 68-77 (1981)
%
 \bibitem{Contopoulos82}
  G. Contopoulos, P. Magnenat, and L. Martinet  {\it Physica D} 
  {\bf 6}, 123-136 (1982)
%
 \bibitem{Froeschle72}
  C. Froeschl\'{e} {\it Astron. Astrophys.} {\bf 16}, 
 172-189 (1972)
%
 \bibitem{Lange14}
  S. Lange, M. Richter,F. Onken, A. B\"{a}cker and R. Ketzmerick {\it Chaos} 
 {\bf 24}, 024409 (2014)
%
 \bibitem{Richter14}
  M. Richter, S. Lange, A. B\"{a}cker and R. Ketzmerick {\it Phys. Rev. E} 
 {\bf 89}, 022902 (2014)
%
 \bibitem{Patsis94}
  P.~A. Patsis  and L. Zachilas {\it Int. J. Bif. Chaos} {\bf 4}, 
  1399-1424 (1994)
%
 \bibitem{Katsanikas11a}
  M. Katsanikas and  P.A. Patsis   {\it Int. Journal Bif. Chaos}  
  {\bf 21}, 467-496 (2011)
%
 \bibitem{Katsanikas11b} 
  M. Katsanikas, P.A Patsis and G. Contopoulos {\it Int. Journal Bif. Chaos} 
  {\bf 21}, 2321-2330 (2011)
%
 \bibitem{Katsanikas11c}
  M. Katsanikas, P.A. Patsis and A.D. Pinotsis  {\it Int. Journal Bif. Chaos} 
  {\bf 21}, 2331-2342 (2011)
%
 \bibitem{Katsanikas13}
  M. Katsanikas, P.A Patsis and G. Contopoulos {\it Int. Journal Bif. Chaos} 
  {\bf 23}, 1330005 (2013)
%
 \bibitem{Patsis14a}
  P.A. Patsis and M. Katsanikas {\it Mon. Not. R. Astron. Soc.} 
  {\bf 445}, 3525-3545 (2014)
%
 \bibitem{Patsis14b}
  P.A. Patsis and M. Katsanikas  {\it Mon. Not. R. Astron. Soc.} 
  {\bf 445}, 3546-3556 (2014) 
%
 \bibitem{Contopoulos02}
  G. Contopoulos {\it Order and Chaos in Dynamical Astronomy}, Springer-Verlag, 
  New York Berlin Heidelberg (2002)  
%
 \bibitem{Zachilas13}
  L.Zachilas, M. Katsanikas and P.A. Patsis {\it Int. Journal Bif. Chaos}  
  {\bf 23}, 1330023 (2013)
%
 \bibitem{Geisel13}
  Christopher  D. Geisel, {\it Spacecraft Orbit Design in the Circular
  Restricted Three-Body Problem Using Higher-Dimensional Poincar\'{e} Maps},
  PhD Thesis, Purdue University, West Lafayette, Indiana, USA (2013)
%
 \bibitem{LGSK} 
 G. Lukes-Gerakopoulos, J. Seyrich, and D. Kunst {\it Phys. Rev. D}
 {\bf  90}, 104019 (2014)
%
 \bibitem{Skokos97}
 Ch. Skokos, G. Contopoulos G. and C. Polymilis {\it Celest. Mech. 
    Dyn. Astron.}  {\bf 65}, 223-251
%
 \bibitem{Vrahatis96}
  M.N. Vrahatis , T.C. Bountis  and M. Kollmann   
  {\it Int. J. Bif. Chaos} {\bf 6}, 1425-1437 (1996)
%
 \bibitem{Ruangsri15}
 U. Ruangsri, S.~J. Vigeland, and S.~A. Hughes, arXiv:1512.00376 
%
 \bibitem{SemSuk}
 O. Semer\'{a}k, and P. Sukov\'{a}, {\it Mon. Not. R. Astron. Soc.} {\bf 404}, 545
 (2010); {\bf 425}, 2455 (2012); P. Sukov\'{a}, and O. Semer\'{a}k, 
 {\it Mon. Not. R. Astron. Soc.} {\bf 436}, 978 (2013); V. Witzany, 
 O. Semer\'{a}k, and P. Sukov\'{a}, {\it Mon. Not. R. Astron. Soc.}
 {\bf 451}, 1770 (2015)  
%
  \bibitem{nonKerr}
 G. Lukes-Gerakopoulos, T.~A. Apostolatos and G. Contopoulos, {\it Phys. Rev. D}
 {\bf 81}, 124005 (2010); G. Contopoulos, G. Lukes-Gerakopoulos and T.~A. Apostolatos,
 {\it Int. J. Bifurc. Chaos} {\bf 21}, 2261 (2011) 
 
 
\end{thebibliography}
\end{document}